\documentclass[prl,aps,preprint,showpacs,superscriptaddress]{revtex4}

\usepackage{graphicx}

\begin{document}

  \title{Origin of the Ising Ferrimagnetism 
    and Spin-Charge Coupling
  in LuFe$_2$O$_4$}

  \author{H. J. Xiang}
  \affiliation{National Renewable Energy Laboratory, Golden, Colorado
    80401, USA}
  
  \author{E. J. Kan}
  \affiliation{Department of Chemistry, North Carolina State
    University, Raleigh, North Carolina 27695-8204, USA}

  \author{Su-Huai Wei}
  \affiliation{National Renewable Energy Laboratory, Golden, Colorado 80401, USA}

  \author{M.-H. Whangbo}
  \affiliation{Department of Chemistry, North Carolina State
    University, Raleigh, North Carolina 27695-8204, USA}

  \author{Jinlong Yang}
  \affiliation{Hefei National Laboratory for Physical Sciences at Microscale,
    University of Science and Technology of China, Hefei, Anhui 230026,
    P. R. China}

  \date{\today}

  \begin{abstract}  
    The spin ordering and spin-charge coupling in LuFe$_2$O$_4$
    were investigated on the basis of density functional calculations
    and Monte Carlo simulations.
    The 2:1 ferrimagnetism arises from the strong antiferromagnetic
    intra-sheet Fe$^{3+}$-Fe$^{3+}$ and Fe$^{3+}$-Fe$^{2+}$
    as well as some substantial antiferromagnetic Fe$^{2+}$-Fe$^{3+}$
    inter-sheet 
    spin exchange interactions.
    The giant magnetocapacitance at room temperature and the enhanced 
    electric polarization at 240 K of LuFe$_2$O$_4$ are 
    explained by the strong spin-charge coupling.        
  \end{abstract}

  \pacs{75.80.+q,71.20.-b,77.80.-e,64.60.De}

  \maketitle

\begin{figure}
    \includegraphics[width=7.5cm]{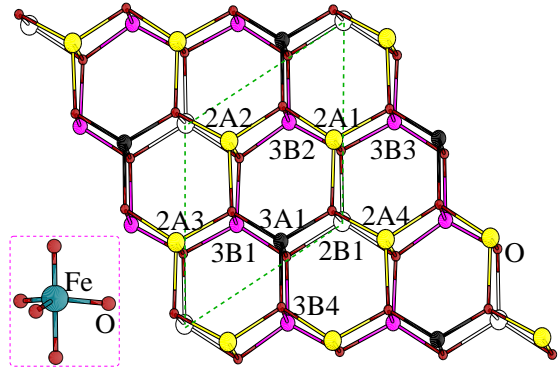}
    \caption{(Color online) Schematic representation of the $\sqrt{3}\times \sqrt{3}$
      CO structure. Large, medium, and small circles
      represent the Fe$^{2+}$, Fe$^{3+}$, and O$^{2-}$ ions,
      respectively. The type A (type B) T-sheet has the honeycomb network of 
      Fe$^{2+}$ (Fe$^{3+}$) ions with a Fe$^{3+}$ (Fe$^{2+}$) ion at the center 
      of each hexagon. 2A and 3A (2B and 3B) refer to the Fe$^{2+}$ and Fe$^{3+}$ 
      ions of the type A (type B) T-sheet, respetively. 
      The region enclosed by dashed lines indicates the unit cell of 
      the CO structure. There is a mirror plane of symmetry, which is parallel to
      the $c$ axis and crosses the 3A1 and 2B1 sites. The inset shows
      an isolated FeO$_5$ trigonal bipyramid.} 
    \label{fig1}
  \end{figure}

  Recently, multiferroics \cite{Kimura2003, Hur2004, Ikeda2005,Xu2008,Subramanian2006,
  Zhang2008,Xiang2007A,Xiang2007B,Xiang2008,Angst2008} have attracted much attention
  because of their potential applications in novel magnetoelectric and
  magneto-optical devices. Among the newly discovered multiferroics,
  LuFe$_2$O$_4$ is particularly interesting due to its
  large ferroelectric (FE) polarization \cite{Ikeda2005} 
  and giant magnetocapacitance at room temperature \cite{Subramanian2006}. 
  In the high-temperature crystal structure of LuFe$_2$O$_4$ with space group R\=3m, 
  layers of composition Fe$_2$O$_4$ alternate with layers of Lu$^{3+}$ ions, 
  such that there are
  three Fe$_2$O$_4$ layers per unit cell. Each Fe$_2$O$_4$ layer
  is made up of two triangular sheets (hereafter, T-sheets) of
  corner-sharing FeO$_5$ trigonal bipyramids (Fig.~\ref{fig1}).
  Below 320 K ($T_{CO}$) LuFe$_2$O$_4$ undergoes a three-dimensional (3D) 
  charge ordering (CO) (2Fe$^{2.5+}$ $\Rightarrow$ Fe$^{2+}$ + Fe$^{3+}$) with the 
  $\sqrt{3}\times \sqrt{3}$ superstructure in each T-sheet;  
  in each Fe$_2$O$_4$ layer, 
  one T-sheet has the honeycomb network of Fe$^{2+}$ 
  ions with a Fe$^{3+}$ ion at the center of each Fe$^{2+}$ hexagon
  (hereafter, the type A  T-sheet), 
  while the other T-sheet has an opposite arrangement 
  of the  Fe$^{2+}$ and Fe$^{3+}$ ions
  (hereafter the type B T-sheet).

  LuFe$_2$O$_4$, with the novel CO-driven ``electronic ferroelectricity'',  
  \cite{Ikeda2005} presents
  several fundamental questions. 
  First, LuFe$_2$O$_4$ shows strong  Ising behavior with the
  easy axis along $c$ \cite{Iida1993,Wu2008}. The spin anisotropy 
  of the non-CO state is
  understandable because the spin down electron of the Fe$^{2.5+}$ ion
  partially occupies the degenerate ($d_{x^2-y^2}$,$d_{xy}$) orbitals
  \cite{Xiang2007A,Dai2005}.
  However, the Ising behavior 
  below $T_{CO}$ is puzzling because the insulating $\sqrt{3}\times \sqrt{3}$ CO
  breaks the 3-fold
  rotational symmetry hence lifting the degeneracy of the 
  ($d_{x^2-y^2}$,$d_{xy}$) orbitals
  \cite{Xiang2007A}.
  Second, LuFe$_2$O$_4$ undergoes a ferrimagnetic spin ordering below 240 K ($T_N$)
  \cite{Iida1993, Tanaka1989, Siratori1992, Christianson2008}. 
  A number of experimental studies found this spin ordering to be
  two-dimensional (2D) in nature \cite{Iida1993,
  Tanaka1989,Funagashi1984}. In contrast, a recent neutron diffraction 
  study observed a finite spin correlation along $c$ and suggested a
  3D spin structure without considering 
  CO \cite{Christianson2008}.
  The M\"ossbauer \cite{Tanaka1989} and neutron diffraction \cite{Siratori1992}
  studies led to a detailed ferrimagnetic structure of LuFe$_2$O$_4$, 
  in which the majority spin lattice consists of all Fe$^{2+}$ ions plus 
  one-third of the total Fe$^{3+}$ ions 
  while the minority spin sublattice consists of the remaining Fe$^{3+}$ ions.
  This 2:1 ferrimagnetic order was suggested to originate from weak 
  ferromagnetic (FM)
  interactions between the next-nearest neighbor (NNN) Fe sites in the triangular
  antiferromagnetic (AFM) Ising lattice \cite{Iida1993}.
  However, 
  using the spin exchange parameters estimated from the
  energy parameters of LaFeO$_3$, 
  Naka {\it et al.} \cite{Naka2008} predicted quite a different
  spin structure that includes some Fe sites without unique spin direction.
  Therefore, the detailed ferrimagnetic
  structure and its origin remain unclear.
  Third, LuFe$_2$O$_4$ exhibits a giant magnetodielectric
  response at room temperature \cite{Subramanian2006}, and a room-temperature
  dynamic magnetoelectric coupling was also reported \cite{Park2007}.
  Furthermore, the FE polarization of LuFe$_2$O$_4$ was found to increase 
  around $T_{N}$ \cite{Ikeda2005}. These observations suggest 
  the occurrence of coupling between the CO and magnetism. 
  The understanding of the spin-charge 
  coupling is crucial for future magnetodielectric applications of LuFe$_2$O$_4$.    
  
  In this Letter, we explore these isuues on the basis of first principles density 
  functional calculations for the first time. 
  A large spin anisotropy is found along the $c$ direction due mainly to 
  the Fe$^{2+}$ ions of the B-sheet, the spin ground state of the
  $\sqrt{3}\times \sqrt{3}$ CO state has the 2:1 
  ferrimagnetic spin arrangement proposed by Siratori {\it et al.}
  \cite{Siratori1992},
  and there occurs strong spin-charge coupling in LuFe$_2$O$_4$. 
  
  Our density functional theory calculations employed the frozen-core projector
  augmented wave method \cite{PAW} encoded in the Vienna {\it ab initio}
  simulation package \cite{VASP}, and the generalized-gradient
  approximation (GGA) \cite{Perdew1996}.
  To properly describe the strong electron correlation in the 3d
  transition-metal oxide, the GGA plus on-site repulsion U method
  (GGA+U) \cite{Liechtenstein1995} was employed with the effective $U$ value
  ($U_{eff} = U -J$ with $J = 0 $) of 4.61 eV \cite{Xiang2007A}. 
  It is known experimentally \cite{Iida1993,Tanaka1989,Funagashi1984} 
  that the interlayer magnetic interactions in LuFe$_2$O$_4$ are weak,
  which is understandable due to its layered structure.
  In this work, therefore, we focus on the 2D spin ordering 
  within a single Fe$_2$O$_4$ layer. For the 
  $\sqrt{3}\times \sqrt{3}$ CO state of LuFe$_2$O$_4$, the FE ordering of 
  the Fe$_2$O$_4$ layers will be assumed.     
  
  We first examine the magnetic anisotropy of the Fe
  ions by performing GGA+U calculations, with spin-orbit
  coupling (SOC) included, for the FM state of LuFe$_2$O$_4$ 
  with the $\sqrt{3}\times \sqrt{3}$ CO. 
  As shown in Fig.~\ref{fig1}(a), there are two 
  kinds of Fe$^{2+}$ ions and two kinds of Fe$^{3+}$ ions in the
  $\sqrt{3}\times \sqrt{3}$ CO state. 
  We label the Fe$^{2+}$ and Fe$^{3+}$ ions of the type A T-sheet 
  as 2A and 3A, respectively, and those of the type B T-sheet 
  as 2B and 3B, respectively. In our GGA+U+SOC calculations
  with spins pointing along several different directions,
  all Fe$^{2+}$ and Fe$^{3+}$ spins are kept
  in the same direction. Our calculations show that 
  the easy axis is along the $c$ direction, as experimentally observed
  \cite{Iida1993,Wu2008}; the $\parallel$c-spin orientation is more stable than the 
  $\perp$c-spin orientation by 1.5 meV per formula unit (FU). 
  The orbital moments of 2A, 2B, 3A, and 3B for 
  the $\parallel$c-spin orientation are 
  0.101, 0.156, 0.031 and 0.035, respectively, which are 
  greater than those for the $\perp$c-spin orientation 
  by 0.019, 0.062, 0.015, and 0.018 $\mu_B$, respectively.
  As expected, the Fe$^{3+}$ ($d^5$) ions have a very small anisotropy, 
  However, two kinds of the Fe$^{2+}$ ions also
  have different degree of spin anisotropy. 
  The spin down electron of
  the 2B Fe$^{2+}$ ion occupies the ($d_{x^2-y^2}$,$d_{xy}$) manifold
  \cite{Xiang2007A},
  therefore the 2B Fe$^{2+}$ ion has the largest spin anisotropy along
  $c$. Our calculations indicate a non-negligible orbital
  contribution to the total
  magnetization, in agreement with the   
  X-ray magnetic circular dichroism result \cite{Wu2008}.

  To determine the magnetic ground state of LuFe$_2$O$_4$ in the
  $\sqrt{3}\times \sqrt{3}$ CO state, we extract its spin exchange
  parameters by mapping the energy differences between ordered spin states
  obtained from GGA+U calculations onto the corresponding energy differences 
  obtained from the Ising Hamiltonian \cite{whangbo2003}:
  \begin{equation}
    H=\sum_{i,j}J_{ij} S_{iz} S_{jz},
  \end{equation}
  where the energy is expressed with respect to the spin disorder (paramagnetic)
  state,  $J_{ij}$ is the spin exchange parameter between the spin
  sites $i$ and $j$,  
  and $S_{iz}$ is the spin component along the $c$ direction
  ($|S_z|=2$  and $2.5$  for Fe$^{2+}$ and Fe$^{3+}$ ions, respectively). 
  We consider all 15 possible superexchange (SE) interactions and all 19 
  super-superexchange (SSE) interactions with the O...O distance less than 3.2 \AA. The
  intra- and inter-sheet interactions within each Fe$_2$O$_4$ layer 
  as well as the SSE interactions between adjacent Fe$_2$O$_4$ layers
  are taken into account.
  To evaluate these 34 spin exchange parameters reliably, 
  we considered 111 different ordered spin states leading to 110
  energy differences. The 34 spin exchange parameters were determined by
  performing a linear least-square fitting analysis. 
  The SSE interactions are generally much
  weaker than the SE interactions with the magnitude of all
  SSE  interactions less than 1.4 meV.
  The calculated SE parameters are reported in Table~\ref{table1}.
  All intra-sheet SE interactions are AFM, and the strongest interactions 
  ($\sim 7.3$ meV) occurs between the 3B Fe$^{3+}$ ions because of  the large
  energy gain of the AFM configuration and almost
  zero FM coupling. 
  The inter-sheet SE interactions are weaker than the
  the intra-sheet SE interactions, and are mostly AFM.

\begin{figure}
    \includegraphics[width=7.5cm]{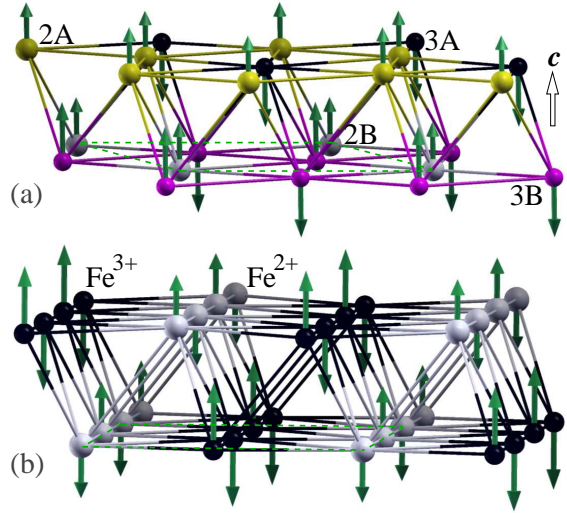}
    \caption{(Color online) Schematic representations of
    (a) the spin ground state of the $\sqrt{3}\times
    \sqrt{3}$ CO structure and (b) one of the macroscopic spin ground states of
    the chain CO structure. The arrows denote the spin
    directions. 
    The region enclosed by the dashed lines on the bottom T-sheet
    indicates the magnetic unit cell of the spin structure.}
    \label{fig2}
  \end{figure}

  With the calculated spin exchange parameters, one can identify the
  spin ground state of the CO state. The Metropolis Monte Carlo
  simulation of the Ising model is performed to search for the ground
  state. Simulations with supercells of several different sizes show that
  the spin ground state has the magnetic structure shown in Fig.~\ref{fig2}(a),
  which has the same cell as the $\sqrt{3}\times \sqrt{3}$ CO structure. 
  In this state, all
  Fe$^{2+}$ ions contribute to the majority spin, and the Fe$^{3+}$ ions 
  are antiferromagnetically coupled to the Fe$^{2+}$ ions in the type A 
  T-sheet. In the honeycomb lattice of the type B T-sheet, 
  the Fe$^{3+}$ spins are antiferromagnetically coupled. Thus, 
  the spin ground state is
  ferrimagnetic, as experimentally observed \cite{Iida1993}.
  This 2:1 ferrimagnetic structure is the same as the magnetic
  structure proposed by Siratori {\it et al.} \cite{Siratori1992}, 
  and differs from the structure 
  proposed by Naka {\it et al.} \cite{Naka2008}. 
       
  The observed ferrimagnetic ordering can be readily explained 
  in terms of the calculated exchange parameters. 
  In the honeycomb network of the type B T-sheet, 
  the nearest-neighbor (NN) 3B ions are antiferromagnetically 
  coupled since their SE interaction is strongly AFM. 
  In the type A T-sheet, the SE interactions 
  between the 2A ions are AFM, and so are 
  those between the 2A and 3A ions, which leads to  
  spin frustration. As a consequence, two possible spin arrangements 
  compete with each other in the type A T-sheet; the first is the state
  in which the coupling between the NN 2A ions are AFM 
  with the spin direction of the 3A ion undetermined, 
  and the second is the state in which all 2A ions are
  antiferromagnetically coupled to the 3A ions. 
  The energies of these two states (considering only the SE interaction) are
  $E_1=-4(J_{2A1,2A2}+J_{2A1,2A4})$ per 3A ion, and
  $E_2=-10(J_{3A1,2A1}+J_{3A1,2A2}+J_{3A1,2A3})+4(J_{2A1,2A2}+J_{2A1,2A4})$ 
  per 3A ion, respectively.
  Due to the relatively strong AFM interactions between the 3A and 2A ions (See
  Table~\ref{table1}) and the large spin of the 3A ions, the second
  state has a lower energy, i.e.,  $E_2$ $<$ $E_1$. 
  Without loss of generality, we can assume
  the 2A (3A) ions constitute the majority (minority) spin 
  in the second state. 
  Now, we examine the spin orientation of the Fe$^{2+}$ ions
  in the type B T-sheet. The intra-sheet interactions of the
  2B ion with 3B ions vanish due to the AFM
  ordering of the 3B ions. As for the inter-sheet
  interactions involving the 2B ions, 
  the dominant one is the AFM interaction of the 2B ion with
  the 3A ion ($J_{3A1-2B1}$ in Table~\ref{table1}). 
  Consequently, we obtain the ferrimangetic ground state shown in Fig.~\ref{fig2}(a),
  in which the spin of the 2B ion contributes to the majority spin of the Fe$_2$O$_4$ layer. 
  For the stability of the ferrimangetic ground state, the 
  inter-sheet interaction is essential. This was neglected in the model Hamiltonian 
  study of Naka {\it et al.} \cite{Naka2008}. The ferrimangetic state
  is not due to the FM interactions between NNN Fe ions of the T-sheet
  because they must be vanishingly weak and mostly AFM.    
      
\begin{figure}
    \includegraphics[width=7.5cm]{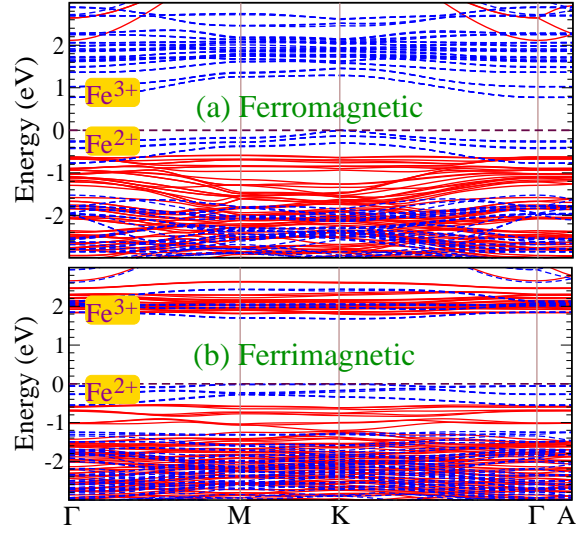}
    \caption{(Color online) Band structures calculated for (a) the FM
    state and (b) the ferrimagnetic state 
    of the $\sqrt{3}\times \sqrt{3}$ CO structure of
    LuFe$_2$O$_4$. The solid and dashed lines represent the up-spin and
    down-spin bands, respectively. The $\sqrt{3}\times \sqrt{3} \times 1$ hexagonal
    cell is used in the calculations. }
    \label{fig3}
  \end{figure}

  The electronic structure of the ferrimangetic state calculated for the
  $\sqrt{3}\times \sqrt{3}$ CO structure of LuFe$_2$O$_4$ is shown in
  Fig.~\ref{fig3}. Also shown is the electronic structure calculated for 
  the FM state. Both states are semiconducting, and the highest
  occupied (HO) and the lowest unoccupied (LU) levels of both states 
  come from the spin-up Fe$^{2+}$
  and Fe$^{3+}$ ions, respectively \cite{Xiang2007A}.
  In addition, the
  band dispersion from $\Gamma$ to A is rather small, indicating a very
  weak interlayer interaction. However, there are some important differences. 
  First, the ferrimangetic state has a larger band gap
  (1.68 eV) than does the FM state (0.77 eV). This is consistent with the
  stability of the ferrimangetic state. Second, the FM state has an
  indirect band gap with the HO and LU levels located at K and $\Gamma$,
  respectively. In the ferrimangetic state, however, the LU level 
  has the highest energy at $\Gamma$ 
  and the band dispersions of the 
  HO and LU levels are almost flat from M to K. This difference comes
  from the orbital interaction between the spin down
  ($d_{x^2-y^2}$,$d_{xy}$) levels of the spin up Fe$^{3+}$ and Fe$^{2+}$
  ions. 

  To probe the presence of spin-charge coupling in LuFe$_2$O$_4$, 
  it is necessary to consider the spin ordering in a CO state other than the
  $\sqrt{3}\times \sqrt{3}$ CO state. 
  The previous electrostatic calculations \cite{Xiang2007A,Naka2008} showed 
  that the chain CO, in which one-dimensional (1D) chains of Fe$^{2+}$ ions 
  alternate with 1D chains of Fe$^{3+}$ ions in each T-sheet 
  [Fig.~\ref{fig2}(b)], is only slightly less stable than the 
  $\sqrt{3}\times \sqrt{3}$ CO, and has no FE polarization.  
  We extract exchange parameters by mapping analysis as described above.
  It is found that the intra-sheet SE between the Fe$^{3+}$ ions is
  the strongest ($J=6.7$ meV) as in the $\sqrt{3}\times \sqrt{3}$ CO case. 
  All intra-sheet SE's are AFM with
  $J$(Fe$^{3+}$-Fe$^{3+}$) $>$ $J$(Fe$^{2+}$-Fe$^{3+}$) $>$ $J$(Fe$^{2+}$-Fe$^{2+}$). 
  The inter-sheet SE between the Fe$^{3+}$ ions
  is very weak ($|J|<0.3$ meV), and that between the
  Fe$^{2+}$ and Fe$^{3+}$ ions is FM with $J=-1.4$ meV. 
  Interestingly, 
  the inter-sheet SE between the Fe$^{2+}$ ions is rather strongly AFM ($J=6.3$ meV). 
  Monte Carlo simulations using these spin exchange parameters 
  indicate that the spin state shown in
  Fig.~\ref{fig2}(b) is the spin ground state. 
  In this spin ordering, the spins within each chain of Fe$^{2+}$ ions or Fe$^{3+}$
  ions are antiferromagnetically coupled. The NN chains of Fe$^{2+}$ ions 
  belonging to different T-sheets are coupled antiferromagnetically, whereas 
  the corresponding chains of Fe$^{3+}$ are almost decoupled.

  The above results show that the spin ordering of  the chain
  CO state is dramatically different from that of
  the $\sqrt{3}\times \sqrt{3}$ CO state.
  The most important difference is that the total spin
  moments are 2.33 $\mu_B$/FU for the $\sqrt{3}\times
  \sqrt{3}$ CO, but 0 $\mu_B$/FU for the chain CO. This evidences a
  strong spin-charge coupling in LuFe$_2$O$_4$.
  The external magnetic field will have different effects on the two
  CO states due to the the Zeeman effect. It is expected that the
  magnetic field will further stabilize the ferrimagnetic  $\sqrt{3}\times
  \sqrt{3}$ CO state. Consequently, an external magnetic
  field will reduce the extent of charge fluctuation and
  hence decrease the dielectric constant. This supports our explanation 
  for the giant magnetocapacitance effect of LuFe$_2$O$_4$
  at room temperature \cite{Xiang2007A} .

  Without considering the inter-sheet interactions, Naka {\it et al.} \cite{Naka2008}
  suggested that the degeneracy of the spin ground state of the
  $\sqrt{3}\times \sqrt{3}$ CO state is of the order O($2^{N/3}$)( N
  is the number of the spin sites), which is much larger than the spin
  degeneracy [O($2^{\sqrt{N}}$)] of the chain CO state. Thus,
  they proposed that spin frustration induces
  reinforcement of the polar  $\sqrt{3}\times \sqrt{3}$  CO by a gain
  of spin entropy. However, our calculations show that 
  there are substantial inter-sheet spin exchange interactions 
  between the 2B1 and 3A1 ions, which would remove the macroscopic
  degeneracy of the spin ground state of the $\sqrt{3}\times \sqrt{3}$
  CO state. The macroscopic degeneracy still persists for the chain
  CO state. Thus, our work provides a picture 
  opposite to what Naka {\it et al.} proposed. 
  Furthermore, we find that the 
  $\sqrt{3}\times \sqrt{3}$  CO state is more favorable for the spin
  ordering than is the chain CO state; 
  with respect to the paramagnetic state, 
  the spin ground state is lower in energy 
  by $-78$ meV/FU for the $\sqrt{3}\times \sqrt{3}$ CO, but by $-57$ meV/FU  
  for the chain CO.  The model of Naka {\it et al.} 
  \cite{Naka2008} predicts that the polar $\sqrt{3}\times \sqrt{3}$  CO state is
  destabilized and the electric polarization is reduced by the
  magnetic field, since it will lift the macroscopic spin degeneracy.
  In contrast, our work predicts that the magnetic field 
  stabilizes the ferrimagnetic
  $\sqrt{3}\times \sqrt{3}$  CO state due to the Zeeman effect, and  
  provides an explanation for why 
  the electric polarization increases when the temperature is lowered
  below the 
  Neel 
  temperature \cite{Ikeda2005}, because the charge
  fluctuation has an onset well below $T_{CO}$ \cite{Xu2008}.

  In summary, our first principles results explain the
  experimentally observed Ising ferrimagnetism, and manifest the
  spin-charge coupling and magnetoelectric effect in LuFe$_2$O$_4$.
  
    
  Work at NREL was supported by the U.S. Department of Energy, under Contract No. DE-AC36-08GO28308, 
  and work at NCSU by the U. S. Department of Energy, under Grant DE-FG02-86ER45259.

  \clearpage

  \clearpage
  \begin{table}
    \caption{Calculated superexchange parameters (in meV) in the
      $\sqrt{3}\times \sqrt{3}$ CO state of LuFe$_2$O$_4$ 
      (For the spin sites of the 2A, 3A, 2B and 3B ions,see
      Fig.~\ref{fig1} )}
    \begin{tabular}{cccccc}
      \hline
      \hline
      A-A &$J_{3A1,2A1}$&$J_{3A1,2A2}$& $J_{3A1,2A3}$  &$J_{2A1,2A2}$ &
      $J_{2A1,2A4}$ \\
      &3.2 & 4.0 & 4.7 & 1.9 & 3.6 \\
      \hline
      B-B &$J_{3B1,3B2}$& $J_{3B1,3B4}$ &
      $J_{2B1,3B1}$&$J_{2B1,3B2}$&$J_{2B1,3B3}$ \\
      &7.0 & 7.6&1.5 & 2.8 & 1.3\\
      \hline
      A-B&$J_{3A1,3B1}$&$J_{3A1,2B1}$&$J_{2A1,2B1}$&$J_{2A1,3B2}$ &
      $J_{2A1,3B3}$ \\
      & 2.0 & 1.9 & $\sim 0$ & $-0.6$  & 1.2 \\
      \hline
      \hline
    \end{tabular}
    \label{table1}
  \end{table}

  \clearpage

\end{document}